\documentclass[fontsize=9pt,DIV=calc,a4paper,twocolumn]{scrartcl}

\usepackage[T1]{fontenc}
\usepackage[table]{xcolor}
\usepackage[format=plain]{caption}
\usepackage{microtype,amsmath,amssymb,mathtools,booktabs,array,xtab,enumitem}
\usepackage[a4paper,left=20mm,right=20mm,top=15mm,bottom=25mm]{geometry}
\usepackage[auth-lg,affil-it]{authblk}
\usepackage[sort&compress,numbers]{natbib}
\usepackage[hidelinks]{hyperref}
\setlist{noitemsep,topsep=0pt,parsep=0pt,partopsep=0pt,leftmargin=*}

\setkomafont{disposition}{\rmfamily\bfseries}
\setkomafont{subsection}{\normalsize}
\RedeclareSectionCommand[afterskip=-.5em]{subsection}

\usepackage[framed,numbered]{matlab-prettifier}
\lstMakeShortInline[style=Matlab-editor]!
\newenvironment{Figure}{\par\medskip\noindent\minipage{\linewidth}}{\endminipage\par\medskip}
\newcolumntype{A}{ >{$} r <{$} @{} >{${}} l <{$} }

\newcommand*\diff{\mathop{}\!\mathrm{d}} 

\title{Free open reference implementation\\of a two-phase PEM fuel cell model}

\begin{document}

\author{Roman Vetter\thanks{} }
\author{J\"{u}rgen O.~Schumacher}
\affil{Institute of Computational Physics (ICP),\\Zurich University of Applied Sciences (ZHAW),\\Wildbachstrasse 21, CH-8401 Winterthur, Switzerland}

\twocolumn[
\begin{@twocolumnfalse}
\maketitle
\begin{abstract}
In almost 30 years of PEM fuel cell modeling, countless numerical models have been developed in science and industrial applications, almost none of which have been fully disclosed to the public. There is a large need for standardization and establishing a common ground not only in experimental characterization of fuel cells, but also in the development of simulation codes, to prevent each research group from having to start anew from scratch. Here, we publish the first open standalone implementation of a full-blown, steady-state, non-isothermal two-phase model for low-temperature PEM fuel cells. It is based on macro-homogeneous modeling approaches and implements the most essential through-plane transport processes in a five-layer MEA. The focus is on code simplicity and compactness with only a few hundred lines of clearly readable code, providing a starting point for more complex model development. The model is implemented as a standalone MATLAB function, based on MATLAB's standard boundary value problem solver. The default simulation setup reflects wide-spread commercially available MEA materials. Operating conditions recommended for automotive applications by the European Commission are used to establish new fuel cell simulation base data, making our program a valuable candidate for model comparison, validation and benchmarking.
\end{abstract}
\vskip\baselineskip
\end{@twocolumnfalse}
]
{
  \renewcommand{\thefootnote}%
    {\fnsymbol{footnote}}
  \footnotetext[1]{Corresponding Author: roman.vetter@zhaw.ch}
}

\section{Introduction}

The development of macro-homogeneous models of the membrane electrode assembly (MEA) of low-temperature proton exchange membrane fuel cells (LT-PEMFCs) goes back almost 30 years, to Springer et al.~\cite{springer:91} and Bernardi \& Verbrugge \cite{bernardi:90,bernardi:91,bernardi:92}. The first non-isothermal variant was published by Fuller \& Newman \cite{fuller:93}. Ever since these pioneering efforts, research and development on numerical simulations of the various transport processes within the MEA have brought forward a large variety of fuel cell models at different length scales, helping scientists and engineers to better understand the complex nonlinear behavior of these promising energy converters.

Even though many of those models are based on the same core functionality, the policy of publishing the mathematical model description but keeping the numerical implementation tightly closed, has forced software developers to reinvent the wheel by starting from scratch over and over again. The unavailability of a fully transparent and easy-to-understand reference implementation of a basic MEA model with spatial resolution is slowing down the advent of modeling in the fuel cell community, some participants of which even hesitate to include modeling in their work altogether. In their comprehensive review article \cite{weber:14}, Weber et al.\ note that ``the majority of PEFC models to date have either been implemented in commercial software such as FLUENT, COMSOL, STAR CD, or implemented in-house. In either case, the source code has not been made available to the public. This has several major drawbacks including (i) lack of validation and comparison between models, (ii) lack of extension capabilities, and (iii) implementation limitations''. They also comment that ``the key disadvantages of most open-source codes are no graphical user interface and a necessary knowledge of Linux OS and [...] C++ or python''.

Open-source code development and validation activities for fuel cells are only recently picking up some steam. The International Energy Agency has launched an annex on open-source modeling of fuel cells systems, but the focus has primarily been on solid oxide fuel cells so far \cite{le:15,beale:16}. To this day, there are only two known open-source codes capable of simulating the state of the art in PEMFC modeling at the cell scale:
\begin{itemize}
\item \textit{OpenFCST} \cite{secanell:14}, a C++ package based on the deal.II finite element library, freely available under the MIT license. It is highly capable, but with more than 120\,000 lines of C++ code as of version 0.3, it is also very heavy-weight and difficult to handle.
\item \textit{FAST-FC} \cite{fast-fc}, a finite volume tool built on top of OpenFOAM. It consists of about 12\,000 lines of code (not counting OpenFOAM) that are published under the GNU General Public License v3, which can pose an insurmountable legal barrier for commercial use.
\end{itemize}
A third open-source toolkit that has been used to study porous media of PEMFCs is \textit{OpenPNM} \cite{gostick:16}, a pore-network model implementation in Python/SciPy. It is freely available under the MIT license and consists of about 25\,000 lines of code.

\begin{figure*}
	\centering
	\includegraphics{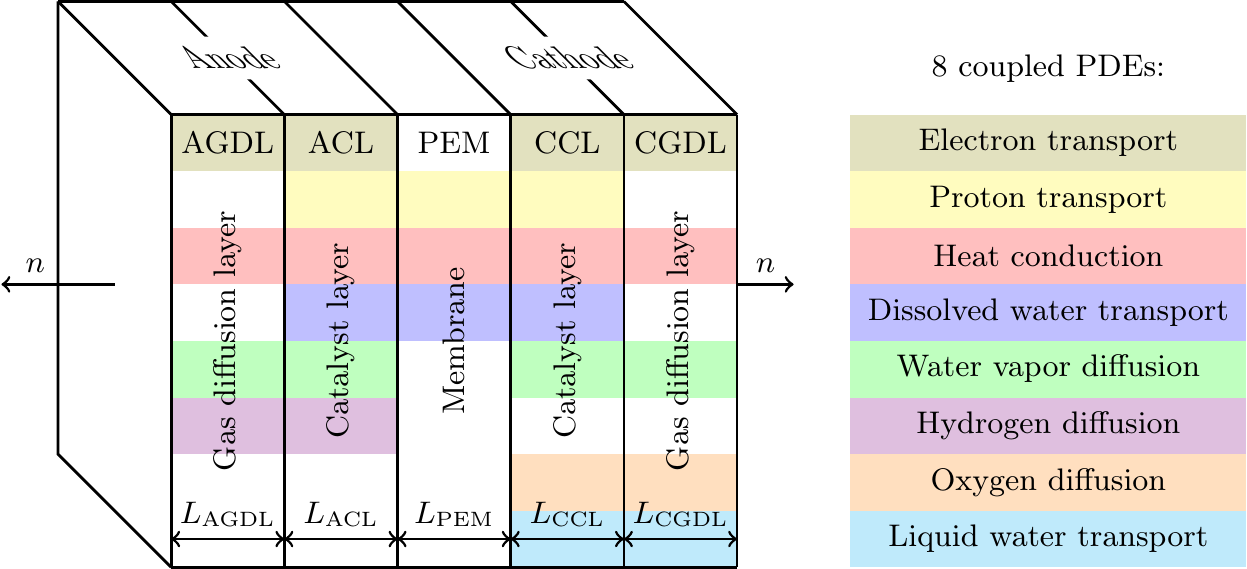}
	\caption{Idealized geometry of the five-layer MEA model (not to scale). Different physical through-plane transport processes are taken into account in different subdomains as marked in color.}
	\label{fig:domain}
\end{figure*}

With this paper, we present a very light-weight, free standalone implementation of a full-blown macro-homogeneous five-layer MEA model for low-temperature PEM fuel cells. The model is formulated in three dimensions, but implemented only in one spatial dimension to represent the dominating through-plane transport processes. It is non-isothermal and two-phase to capture important thermal effects such as phase-change induced flow, but isobaric and stationary to avoid the computational complexity arising from pressure gradients and unsteady behavior. We designed the program code for
\begin{itemize}
\item \textit{Simplicity and compactness.} It comprises less than 400 lines of commented code and does not require manual differentiation of the equations. This dramatically simplifies modifications such as the substitution of individual parameterizations or boundary conditions, which can be done by replacing a single line of code. No lookup tables or data interpolation are involved. 
\item \textit{Portability and compatibility.} The model is implemented as a standalone MATLAB function, relying only on MATLAB's standard boundary value problem solver. This choice of programming environment lets the simulation run on a large variety of platforms and also lets it benefit from MATLAB's widespread availability in science and industry. 
\item \textit{Transparency.} The model equations and boundary conditions are fully disclosed, and the complete simulation output is shown in the paper, including all potentials, fluxes and the entire polarization curve.
\item \textit{Accessibility.} The program is well documented and ready to be used out of the box. Modifying the code requires only minimal programming knowledge, running it requires none at all. All plots are automatically generated.
\item \textit{Free availability.} Our implementation is open-source and published under a 3-clause BSD license permitting also commercial use. It thus provides a starting point for PEMFC model building in industry and research, and a sound basis for modeling extensions such as time dependence, multi-dimensionality, or advanced material parameterizations.
\item \textit{Establishment of a reference simulation.} The model uses long-established and accepted parameterizations of steady-state through-plane transport processes. Widely used commercial MEA materials are used for the default simulation setup. Operating conditions recommended for automotive applications by the Joint Research Centre of the European Commission \cite{tsotridis:15} are simulated, establishing a new baseline for model comparison, benchmarking and validation.
\end{itemize}

\section{Mathematical model}

In the one-dimensional PEMFC model developed here, the MEA is represented as a series of five adjacent homogeneous interval subdomains representing a cell-area-averaged through-plane section of the porous layers of a fuel cell. It includes the proton exchange membrane (PEM) in the middle, sandwiched by two catalyst layers (CLs) and two gas diffusion layers (GDLs). The gas channels (GCs) and mono- or bipolar plates on either end are modeled as boundaries of the MEA. Other subdomains such as microporous layers (MPLs) are not explicitly modeled, but can be added without difficulty. The geometry of the MEA model is shown in Fig.~\ref{fig:domain}. We keep the mathematical description as compact as possible, briefly summarizing the conservation laws and transport equations of the model.

\begin{table*}
	\centering
	\caption{Governing equations.\strut}
	\label{tab:equations}
	\begin{tabular}{lcAA}
	\toprule
	Name & Dependent variable & \multicolumn{2}{c}{Flux} & \multicolumn{2}{c}{Continuity equation}\\
	\midrule
	Ohm's law for electrons & $\phi_\mathrm{e}$ & j_\mathrm{e}&=-\sigma_\mathrm{e}\nabla\phi_\mathrm{e} & \nabla\cdot j_\mathrm{e}&=S_\mathrm{e}\\
	Ohm's law for protons & $\phi_\mathrm{p}$ & j_\mathrm{p}&=-\sigma_\mathrm{p}\nabla\phi_\mathrm{p} & \nabla\cdot j_\mathrm{p}&=S_\mathrm{p}\\
	Fourier heat conduction & $T$ & j_T&=-k\nabla T & \nabla\cdot j_T&=S_T\\
	Water transport in ionomer & $\lambda$ & j_\lambda&=-(D_\lambda/V_\mathrm{m})\nabla\lambda+(\xi/F)j_\mathrm{p} & \nabla\cdot j_\lambda&=S_\lambda\\
	Fickean water vapor diffusion & $x_{\mathrm{H}_2\mathrm{O}}$ & j_{\mathrm{H}_2\mathrm{O}}&=-CD_{\mathrm{H}_2\mathrm{O}}\nabla x_{\mathrm{H}_2\mathrm{O}} & \nabla\cdot j_{\mathrm{H}_2\mathrm{O}}&=S_{\mathrm{H}_2\mathrm{O}}\\
	Fickean hydrogen diffusion & $x_{\mathrm{H}_2}$ & j_{\mathrm{H}_2}&=-CD_{\mathrm{H}_2}\nabla x_{\mathrm{H}_2} & \nabla\cdot j_{\mathrm{H}_2}&=S_{\mathrm{H}_2}\\
	Fickean oxygen diffusion & $x_{\mathrm{O}_2}$ & j_{\mathrm{O}_2}&=-CD_{\mathrm{O}_2}\nabla x_{\mathrm{O}_2} & \nabla\cdot j_{\mathrm{O}_2}&=S_{\mathrm{O}_2}\\
	Liquid water transport (Darcy) & $s$ & j_s&=-(\kappa/\mu V_\mathrm{w})(\partial p_\mathrm{c}/\partial s)\nabla s & \nabla\cdot j_s&=S_s\\
	\bottomrule
	\end{tabular}
\end{table*}

\subsection*{Electrochemistry.}

In hydrogen-fueled LT-PEMFCs, the net electrochemical reaction is
\begin{equation}
\label{eq:reaction}
\mathrm{H}_2(\mathrm{g})+\frac{1}{2}\mathrm{O}_2(\mathrm{g})\to\mathrm{H}_2\mathrm{O}(\mathrm{l})
\end{equation}
and hence the reversible cell potential is given by the Nernst equation \cite{li:06}
\begin{equation}
\label{eq:nernst}
\Delta\phi_0 = -\frac{\Delta G}{2F}+\frac{RT}{2F}\ln\left[\left(\frac{p_{\mathrm{H}_2}}{P_\mathrm{ref}}\right)\left(\frac{p_{\mathrm{O}_2}}{P_\mathrm{ref}}\right)^{1/2}\right]
\end{equation}
where $F$ is the Faraday constant, $R$ the gas constant, $P_\mathrm{ref}=1\,\mathrm{atm}$ is the reference pressure, $T$ is the temperature, $p_{\mathrm{H}_2}=x_{\mathrm{H}_2}P$ and $p_{\mathrm{O}_2}=x_{\mathrm{O}_2}P$ are the partial pressures of hydrogen and oxygen and $\Delta G=\Delta H-T\Delta S$ is the Gibbs free energy change of the reaction. We now assume that the overall redox reaction can be split into a single-step hydrogen oxidation reaction (HOR) in the ACL and a single-step oxygen reduction reaction (ORR) in the CCL, and that both can be described with sufficient accuracy by Butler--Volmer kinetics. The reaction rate in the homogenized catalyst layers is thus locally given by \cite{hamann:98,li:06,ohayre:09}
\begin{equation}
\label{eq:bv}
i = i_0 a\left(\exp\left[\frac{\beta 2F}{RT}\eta\right] - \exp\left[-\frac{(1-\beta)2F}{RT}\eta\right]\right)
\end{equation}
with (positive) activation overpotential
\begin{equation}
\eta = \begin{dcases*}
\Delta\phi-\Delta\phi_0 & in ACL\\
\Delta\phi_0-\Delta\phi & in CCL
\end{dcases*}.
\end{equation}
$\Delta\phi = \phi_\mathrm{e}-\phi_\mathrm{p}$ is the Galvani potential difference between the electron and proton conducting phases (see Tab.~\ref{tab:equations}), and the reversible potential difference $\Delta\phi_0$ is locally divided into
\begin{equation}
\Delta\phi_0 = \begin{dcases*}
-\frac{\phantom{\Delta H-{}}T\Delta S_\mathrm{HOR}}{2F}-\frac{RT}{2F}\ln\left[\frac{p_{\mathrm{H}_2}}{P_\mathrm{ref}}\right] & in ACL\\
-\frac{\Delta H-T\Delta S_\mathrm{ORR}}{2F}+\frac{RT}{4F}\ln\left[\frac{p_{\mathrm{O}_2}}{P_\mathrm{ref}}\right] & in CCL
\end{dcases*}
\end{equation}
with total reaction entropy $\Delta S=\Delta S_\mathrm{HOR}+\Delta S_\mathrm{ORR}$. The sign convention used here is such that a positive $i$ corresponds to a source of positive charge or mass in the continuity equations (see Tab.~\ref{tab:sources}).

\subsection*{Transport of charge, heat and mass.}

For the remainder of the model description we follow the continuum approach to describe the most dominant transport processes of charge, energy, gas species and water by conservation laws. This results in eight coupled second-order partial differential equations (PDEs), which are summarized in Tab.~\ref{tab:equations} for the steady state.

In the CLs and GDLs, Ohm's law is assumed to govern how the flux of electrons $j_\mathrm{e}$ is driven by a gradient of the electronic phase potential. The analogous equation is used for the flux of protons $j_\mathrm{p}$ within the electrolyte phase of the CLs and the membrane. The two electrostatic phase potentials $\phi_\mathrm{e}$ and $\phi_\mathrm{p}$ coexist in the CLs (see Fig.~\ref{fig:domain}), defining $\Delta\phi$ in these two domains. The approach is adopted from the classical porous-electrode theory of Newman, where it is assumed that the electric double layer constitutes only a small volume compared to any of the phases or the electrode itself \cite{newman:75}.

Heat conduction is the dominating mode of energy transport in the MEA \cite{bhaiya:14}, allowing for an accurate description of the heat flux $j_T$ by Fourier's law in all five subdomains. This is the third differential equation.

The description of water balance in the ionomer is based on the seminal model by Springer et al.~\cite{springer:91}. To represent the degree of humidification, the number of water molecules per acidic group $\lambda$ is used. The molar flux of dissolved water $j_\lambda$ is composed of the sum of its two most significant contributions: back diffusion due to a moisture gradient ($j_\lambda\sim-\nabla\lambda$) and electro-osmotic drag ($j_\lambda\sim j_\mathrm{p}\sim-\nabla\phi_\mathrm{p}$).

The next three equations are dedicated to the transport of gas species on both sides of the membrane. If gas crossover is neglected, it is sufficient to consider hydrogen only on the anode side and oxygen on the cathode side, whereas water vapor is present in both gas mixtures. A third gas component is implicitly accounted for on either side (typically nitrogen if air is supplied to the cathode) and need not explicitly be computed because the sum of mole fractions $\sum_Xx_X=1$ everywhere. We assume uniform gas pressure $P$ in the steady state, and hence the dominant transport mechanism is inter-diffusion of the gas species. The simplest transport model for this is Fick's law $j_X=-CD_X\nabla x_X$ \cite{bird:02}, which is employed here for all species. Thermal diffusion, as it results from the chemical potential gradient as the general driving force for species transport, is neglected. The ideal gas law is used to calculate the interstitial gas concentration $C=P/RT$.

Finally, for the description of liquid water transport, we adopt the unsaturated flow theory that was carried over from soil physics to the fuel cell modeling community by Natarajan \& Nguyen \cite{natarajan:01}, and which has since become the de-facto standard in macro-homogeneous two-phase MEA modeling. Darcy's law is transformed into an equation for liquid water flux driven by a gradient $\nabla s$, where $s$ denotes the liquid water saturation (fraction of pore space filled with liquid water). This requires the specification of both the saturation-dependent hydraulic permeability $\kappa$ and the differential relationship between the capillary pressure and saturation, $\partial p_\mathrm{c}/\partial s$, as material properties.

For each of these eight fluxes, a continuity equation is expressed in the last column of Tab.~\ref{tab:equations}, equating the divergence of each flux $j$ with a corresponding source term $S$. These eight PDEs become nonlinear when the coefficients and/or source terms are expressed in terms of the dependent variables. All phase transitions and reaction rates appear as sources that couple these PDEs as detailed in the following.

\subsection*{Source terms and phase transitions.}

A summary of all source term definitions is given in Tab.~\ref{tab:sources}. In the ACL, hydrogen is split into electrons and protons with a reaction rate given by Eq.~\ref{eq:bv}, giving rise to source terms $S_\mathrm{e}$ and $S_\mathrm{p}$. Faraday's law determines the rate of hydrogen consumption in the ACL as well as the oxygen consumption and water production in the CCL (moles consumed or produced per unit volume, time and exchanged electron pair):
\begin{equation}
S_\mathrm{F} = \frac{i}{2F}
\end{equation}
The molar oxygen consumption rate is only half this much (see Eq.~\ref{eq:reaction}). It is assumed that water is produced at the platinum--ionomer phase boundary in dissolved form \cite{wu:09b}, hence $S_\mathrm{F}$ appears as a contribution to $S_\lambda$ in the CCL.

\begin{table*}
	\centering
	\caption{Source terms.\strut}
	\label{tab:sources}
	\begin{tabular}{lcccccc}
	\toprule
	Source && AGDL & ACL & PEM & CCL & CGDL\\
	\midrule
	$S_\mathrm{e}$ & $=$ & $0$ & $-i$ & & $i$ & $0$\\
	$S_\mathrm{p}$ & $=$ & & $i$ & $0$ & $-i$ & \\
	$S_T$ & $=$ & $S_{T,\mathrm{e}}$ & $S_{T,\mathrm{e}}+S_{T,\mathrm{p}}+S_{T,\mathrm{r}}+S_{T,\mathrm{ad}}$ & $S_{T,\mathrm{p}}$ & $S_{T,\mathrm{e}}+S_{T,\mathrm{p}}+S_{T,\mathrm{r}}+S_{T,\mathrm{ad}}+S_{T,\mathrm{ec}}$ & $S_{T,\mathrm{e}}+S_{T,\mathrm{ec}}$\\
	$S_\lambda$ & $=$ & & $S_\mathrm{ad}$ & $0$ & $S_\mathrm{F}+S_\mathrm{ad}$ & \\
	$S_{\mathrm{H}_2\mathrm{O}}$ & $=$ & $0$ & $-S_\mathrm{ad}$ & & $-S_\mathrm{ad}-S_\mathrm{ec}$ & $-S_\mathrm{ec}$\\
	$S_{\mathrm{H}_2}$ & $=$ & $0$ & $-S_\mathrm{F}$\\
	$S_{\mathrm{O}_2}$ & $=$ & & & & $-S_\mathrm{F}/2$ & $0$\\
	$S_s$ & $=$ & & & & $S_\mathrm{ec}$ & $S_\mathrm{ec}$\\
	\bottomrule
	\end{tabular}
\end{table*}

Absorption and desorption of water vapor into/from the ionomer does not happen instantaneously, but at a finite rate over a time span of the order of an hour \cite{rivin:01,majsztrik:07,satterfield:08,cheah:11}. To account for this significant ionomer--gas interfacial water transport resistance, the sorption source term $S_\mathrm{ad}$ appearing in the continuity equations of $\lambda$ and $x_{\mathrm{H}_2\mathrm{O}}$ is set to \cite{wu:09}
\begin{equation}
\label{eq:Sad}
S_\mathrm{ad} =
\begin{dcases*}
\frac{k_\mathrm{a}}{LV_\mathrm{m}}(\lambda_\mathrm{eq}-\lambda) & if $\lambda < \lambda_\mathrm{eq}$ (absorption)\\
\frac{k_\mathrm{d}}{LV_\mathrm{m}}(\lambda_\mathrm{eq}-\lambda) & if $\lambda > \lambda_\mathrm{eq}$ (desorption)
\end{dcases*}
\end{equation}
where $L$ is the thickness of the CL, $V_\mathrm{m}$ the molar charge volume of the ionomer, $\lambda_\mathrm{eq}$ denotes the RH-dependent equilibrium water content of the ionomer, and $k_\mathrm{a}$, $k_\mathrm{d}$ are material-dependent mass-transfer coefficients.

The commonly employed approach to include liquid--vapor phase change in macro-homogeneous MEA modeling is to assume the mass transfer to be driven by the vapor partial pressure difference to the saturation pressure \cite{fritz:09}. With $x_\mathrm{sat}=P_\mathrm{sat}/P$, the corresponding water source/sink term can be expressed as
\begin{equation}
S_\mathrm{ec} =
\begin{dcases*}
\gamma_\mathrm{e}C(x_{\mathrm{H}_2\mathrm{O}}-x_\mathrm{sat}) & if $x_{\mathrm{H}_2\mathrm{O}} < x_\mathrm{sat}$ (evap.)\\
\gamma_\mathrm{c}C(x_{\mathrm{H}_2\mathrm{O}}-x_\mathrm{sat}) & if $x_{\mathrm{H}_2\mathrm{O}} > x_\mathrm{sat}$ (cond.)
\end{dcases*}
\end{equation}
where $\gamma_\mathrm{e}$ and $\gamma_\mathrm{c}$ are the evaporation and condensation rates.

The latent heat released or absorbed during the above two phase transitions can be modeled by adding the following contributions to the total heat source $S_T$:
\begin{equation}
\begin{aligned}
S_{T,\mathrm{ad}} &= H_\mathrm{ad}S_\mathrm{ad}\\
S_{T,\mathrm{ec}} &= H_\mathrm{ec}S_\mathrm{ec}
\end{aligned}
\end{equation}
$H_\mathrm{ad}$ and $H_\mathrm{ec}$ are the molar enthalpies of desorption and vaporization. Joule's first law provides two more heat sources induced by the electric and ionic currents,
\begin{equation}
\begin{aligned}
S_{T,\mathrm{e}} &= \sigma_\mathrm{e}(\nabla\phi_\mathrm{e})^2 = -j_\mathrm{e}\cdot\nabla\phi_\mathrm{e}\\
S_{T,\mathrm{p}} &= \sigma_\mathrm{p}(\nabla\phi_\mathrm{p})^2 = -j_\mathrm{p}\cdot\nabla\phi_\mathrm{p}.
\end{aligned}
\end{equation}
And finally, the heat dissipated by the electrochemical reaction is given by the sum of activation and Peltier heats \cite{weber:14}:
\begin{equation}
S_{T,\mathrm{r}} = i\eta-S_\mathrm{F}\times\begin{dcases*}
T\Delta S_\mathrm{HOR} & in ACL\\
T\Delta S_\mathrm{ORR} & in CCL
\end{dcases*}
\end{equation}

\subsection*{Boundary conditions.}

To complete the mathematical model description, a set of boundary conditions (BCs) needs to be specified (two for each contiguous support in each of the eight second-order PDEs), which are listed in Tab.~\ref{tab:bc}. The membrane is assumed to be impermeable for all gas species as well as for electrons and liquid water, hence these normal fluxes vanish at the membrane boundaries. Protons and dissolved water on the other hand are bound to the ionomer phase, which implies zero fluxes at the outer surfaces of the catalyst layers. Since the electrostatic potentials can be freely offset, $\phi_\mathrm{e}$ is set to zero at the outer AGDL boundary. At the remaining interior subdomain interfaces, continuity of the potentials and fluxes is assumed as indicated in Tab.~\ref{tab:bc}.

For the remaining outer boundaries, a reasonable choice depends on the scenario to be simulated. We impose the cell voltage $U$ by applying the Dirichlet BC $\phi_\mathrm{e}=U$ at the other end of the MEA, but equivalently one can control the current density $I$ by using a Neumann BC $n\cdot j_\mathrm{e}=I$ instead. We furthermore impose the gas channel temperature, pressure, relative humidity and gas composition via
\begin{equation}
\begin{aligned}
x_{\mathrm{H}_2\mathrm{O}}^\mathrm{A} &= \mathrm{RH}_\mathrm{A}P_\mathrm{sat}(T_\mathrm{A})/P_\mathrm{A}\\
x_{\mathrm{H}_2\mathrm{O}}^\mathrm{C} &= \mathrm{RH}_\mathrm{C}P_\mathrm{sat}(T_\mathrm{C})/P_\mathrm{C}\\
x_{\mathrm{H}_2}^\mathrm{A} &= \alpha_{\mathrm{H}_2}(1-x_{\mathrm{H}_2\mathrm{O}}^\mathrm{A})\\
x_{\mathrm{O}_2}^\mathrm{C} &= \alpha_{\mathrm{O}_2}(1-x_{\mathrm{H}_2\mathrm{O}}^\mathrm{C})\\
\end{aligned}
\end{equation}
where $\alpha_{\mathrm{H}_2}$ ($\alpha_{\mathrm{O}_2}$) is the hydrogen (oxygen) mole fraction in the supplied gas when dry. 

Perhaps the most delicate interface treatment is that of liquid water, and the formulation of physically accurate models is a topic of ongoing research in PEMFC modeling \cite{weber:14}. At the CGDL/GC interface, water droplets form and detach in a dynamic fashion (e.g., \cite{yang:04,schillberg:07}) that is difficult to translate into a steady-state area-averaged BC. Historically, simple Dirichlet BCs for $s$ are often used \cite{natarajan:01,nam:03,lin:04}. Conditional unidirectional flux conditions have later been proposed as a more realistic replacement \cite{weber:06,gerteisen:09}, but solving these numerically can be a challenge \cite{zhou:17}. We use here a Dirichlet BC for $s$ at the CGDL/GC interface---the simplest common denominator in two-phase MEA modeling---bearing its limitations in mind.

\begin{table*}
	\centering
	\caption{Boundary conditions. $n$ denotes the interfacial unit normal vector.\strut}
	\label{tab:bc}
	\begin{tabular}{lcccccc}
	\toprule
	Variable & AGC/AGDL & AGDL/ACL & ACL/PEM & PEM/CCL & CCL/CGDL & CGDL/CGC\\
	\midrule
	$\phi_\mathrm{e}$ & $\phi_\mathrm{e}=0$ & continuity & $n\cdot j_\mathrm{e}=0$ & $n\cdot j_\mathrm{e}=0$ & continuity & $\phi_\mathrm{e}=U$\\
	$\phi_\mathrm{p}$ & & $n\cdot j_\mathrm{p}=0$ & continuity & continuity & $n\cdot j_\mathrm{p}=0$ & \\
	$T$ & $T=T_\mathrm{A}$ & continuity & continuity & continuity & continuity & $T=T_\mathrm{C}$\\
	$\lambda$ & & $n\cdot j_\lambda=0$ & continuity & continuity & $n\cdot j_\lambda=0$ & \\
	$x_{\mathrm{H}_2\mathrm{O}}$ & $x_{\mathrm{H}_2\mathrm{O}}=x_{\mathrm{H}_2\mathrm{O}}^\mathrm{A}$ & continuity & $n\cdot j_{\mathrm{H}_2\mathrm{O}}=0$ & $n\cdot j_{\mathrm{H}_2\mathrm{O}}=0$ & continuity & $x_{\mathrm{H}_2\mathrm{O}}=x_{\mathrm{H}_2\mathrm{O}}^\mathrm{C}$\\
	$x_{\mathrm{H}_2}$ & $x_{\mathrm{H}_2}=x_{\mathrm{H}_2}^\mathrm{A}$ & continuity & $n\cdot j_{\mathrm{H}_2}=0$ & & & \\
	$x_{\mathrm{O}_2}$ & & & & $n\cdot j_{\mathrm{O}_2}=0$ & continuity & $x_{\mathrm{O}_2}=x_{\mathrm{O}_2}^\mathrm{C}$\\
	$s$ & & & & $n\cdot j_\mathrm{s}=0$ & continuity & $s=s_\mathrm{C}$\\
	\bottomrule
	\end{tabular}
\end{table*}

\subsection*{Initial conditions.}

Nonlinear problems require a good initial guess of the solution for iterative solvers to converge. It is most convenient to iterate over cell voltages from high to low to generate the polarization curve, since one can then start with all-zero fluxes as a good initial guess. For the potentials, the following initial conditions are usually sufficient for convergence: $\phi_\mathrm{e} \equiv (0 \mid U)$, $\phi_\mathrm{p} \equiv 0$, $T \equiv (T_\mathrm{A}+T_\mathrm{C})/2$, $\lambda \equiv \lambda_\mathrm{eq}\rvert_{\mathrm{RH}=1}$, $x_{\mathrm{H}_2\mathrm{O}} \equiv (x_{\mathrm{H}_2\mathrm{O}}^\mathrm{A} \mid x_{\mathrm{H}_2\mathrm{O}}^\mathrm{C})$, $x_{\mathrm{H}_2} \equiv x_{\mathrm{H}_2}^\mathrm{A}$, $x_{\mathrm{O}_2} \equiv x_{\mathrm{O}_2}^\mathrm{C}$, $s \equiv s_\mathrm{C}$, where the notation $(A \mid C)$ stands for the two values in the AGDL \& ACL ($A$) and CCL \& CGDL ($C$), respectively.

\section{Parameterization}

For the establishment of a useful reference PEMFC simulation suitable for model comparison and benchmarking, it is important to furnish the model with well-established parameterizations of widely available commercial MEA materials. Nafion NR-211 is the membrane of choice here, owing to the large market share and the vast pool of characterization data of Nafion in the literature \cite{kusoglu:17}. Since Toray carbon paper is among the most comprehensively characterized GDLs in the literature, we use Toray TGP-H-060 material properties to parameterize the GDLs in the model. Together, these materials form a typical modern MEA as it may, for instance, be used for automotive applications. Standard literature data are used for the remaining material-independent electrochemical and physical properties.

\subsection*{Water properties.}

For water produced in liquid form at $25\,^\circ\mathrm{C}$ and $1\,\mathrm{bar}$, the enthalpy of formation is $\Delta H=-285.83\,\mathrm{kJ/mol}$ \cite{chase:98}. The saturation pressure of water vapor $P_\mathrm{sat}$ can be approximated with the Antoine equation in the temperature range $T=50{-}100\,^\circ\mathrm{C}$ \cite{green:08}:
\begin{equation}
\ln\left[\frac{P_\mathrm{sat}}{1\,\mathrm{Pa}}\right] = 23.1963 - \frac{3816.44\,\mathrm{K}}{T-46.13\,\mathrm{K}}
\end{equation}
The same relationship goes by the name of Vogel equation when used for the dynamic viscosity of liquid water $\mu$, and it has the following coefficients in the temperature range $T=2{-}95\,^\circ\mathrm{C}$ \cite{goletz:77}:
\begin{equation}
\ln\left[\frac{\mu}{1\,\mathrm{mPa\,s}}\right] = -3.63148 + \frac{542.05\,\mathrm{K}}{T-144.15\,\mathrm{K}}
\end{equation}
The condensation and evaporation rates may be estimated as \cite{nam:03,wu:09b}
\begin{equation}
\label{eq:gamma_ec}
\begin{aligned}
\gamma_\mathrm{e} &= k_\mathrm{e}a_\mathrm{lg}s_\mathrm{red}\\
\gamma_\mathrm{c} &= k_\mathrm{c}a_\mathrm{lg}(1-s_\mathrm{red})
\end{aligned}
\end{equation}
where $a_\mathrm{lg}\approx2\,\mathrm{m}^2/\mathrm{cm}^3$ is an effective liquid--gas interfacial surface area density scaling factor \cite{wu:09b}, and $k_\mathrm{e}$ and $k_\mathrm{c}$ are the Hertz--Knudsen mass transfer coefficients, given at atmospheric pressure by \cite{marek:01}
\begin{equation}
\begin{rcases*}
k_\mathrm{e}\\
k_\mathrm{c}
\end{rcases*}
= \sqrt{\frac{RT}{2\pi M_\mathrm{w}}}
\times\begin{cases*}
5\!\times\!10^{-4}\\
6\!\times\!10^{-3}
\end{cases*}
\end{equation}
where $M_\mathrm{w}=18\,\mathrm{g}/\mathrm{mol}$ is the molar mass of water. In Eq.~\ref{eq:gamma_ec}, liquid water saturation dependence of the phase change interface is introduced through the reduced saturation
\begin{equation}
s_\mathrm{red} = \frac{s-s_\mathrm{im}}{1-s_\mathrm{im}}
\end{equation}
where $s_\mathrm{im}$ denotes the immobile or inaccessible saturation (i.e., liquid water that does not contribute to transport pathways or phase change, e.g.\ due to spatial isolation). It is estimated as $s_\mathrm{im}=s_\mathrm{C}$. Finally, the latent heat of evaporation/condensation is $H_\mathrm{ec}\approx42\,\mathrm{kJ/mol}$ in the temperature range relevant for PEMFC operation \cite{haynes:16}.

\subsection*{Electrochemical parameters.}

Since the ORR is the rate-limiting half-reaction, it is crucial to model the concentration and temperature dependence of the exchange current density in the cathode with high accuracy. Neyerlin et al.~\cite{neyerlin:06} have obtained the following relationship for a Pt/C cathode:
\begin{multline}
\label{eq:i_0_ORR}
i_0 = 2.45\!\times\!10^{-8}\,\mathrm{A/cm}_\mathrm{Pt}^2\left(\frac{p_{\mathrm{O}_2}}{P_\mathrm{ref}}\right)^{0.54}\\
\times\exp\left[\frac{67\,\mathrm{kJ/mol}}{R}\left(\frac{1}{T_\mathrm{ref}}-\frac{1}{T}\right)\right]
\end{multline}
Albeit measured for a low equivalent weight (EW) ionomer, the activation energy in Eq.~\ref{eq:i_0_ORR} is consistent with reported values for higher EW ionomers (such as 1100 EW Nafion) as well \cite{neyerlin:06}. Furthermore we assume that Eq.~\ref{eq:i_0_ORR}, which was fitted premising the Tafel equation, can be applied to the Butler--Volmer equation without modification. For the much faster HOR, the Butler--Volmer equation has been reported to hold with \cite{neyerlin:07,durst:15}
\begin{equation}
i_0 = 0.27\,\mathrm{A/cm}_\mathrm{Pt}^2\exp\left[\frac{16\,\mathrm{kJ/mol}}{R}\left(\frac{1}{T_\mathrm{ref}}-\frac{1}{T}\right)\right].
\end{equation}
We consider a cathode platinum loading that is three times as high as in the anode: $a=1\!\times\!10^{11}\,\mathrm{cm}_\mathrm{Pt}^2/\mathrm{m}^3$ in the ACL and $a=3\!\times\!10^{11}\,\mathrm{cm}_\mathrm{Pt}^2/\mathrm{m}^3$ in the CCL. The symmetry factor $\beta$ is assumed to be $1/2$ in both half-reactions. Lampinen \& Fomino's values for the half-reaction entropies $\Delta S_\mathrm{HOR} = 0.104\,\mathrm{J/mol\,K}$ and $\Delta S_\mathrm{ORR} = -163.3\,\mathrm{J/mol\,K}$ \cite{lampinen:93} appear to be the most plausible in the literature and are therefore adopted here.

\begin{table*}
	\centering
	\caption{Material and through-plane transport parameters.\strut}
	\label{tab:general_parameters}
	\begin{tabular}{llcccc}
	\toprule
	Symbol & Explanation & Unit & AGDL \& CGDL & ACL \& CCL & PEM\\
	\midrule
	$L$ & Layer thickness & \textmu{}m & $160$ \cite{burheim:11} & $10$ & $25$ \cite{nafion:16}\\
	$\epsilon_\mathrm{i}$ & Ionomer volume fraction & -- & & $0.3$ & $1$\\
	$\epsilon_\mathrm{p}$ & Pore volume fraction & -- & $0.76$ \cite{burheim:11} & $0.4$ \cite{bernardi:92} &\\
	$k$ & Thermal conductivity & W/m\,K & $1.6$ \cite{burheim:11} & $0.27$ \cite{khandelwal:06} & $0.3$ \cite{khandelwal:06}\\
	$\tau$ & Pore tortuosity & -- & $1.6^*$ & $1.6$ \cite{litster:13,babu:16} &\\
	$\kappa_\mathrm{abs}$ & Absolute permeability & m$^2$ & $6.15\!\times\!10^{-12}$ \cite{el-kharouf:12} & $10^{-13}$ \cite{yi:99} &\\
	$\sigma_\mathrm{e}$ & Electrical conductivity & S/m & $1250$ \cite{toray} & $350$ \cite{gode:03} &\\
	\midrule[\heavyrulewidth]
	\multicolumn{6}{l}{$^*$Calculated from $\epsilon_\mathrm{p}/\tau^2\approx0.3$ \cite{chan:12}.}\\
	\end{tabular}
\end{table*}

\subsection*{Ionomer-related parameters.}

For Nafion, the enthalpy of (de-)sorption is almost equal to that of vaporization when not completely dry \cite{wadso:13}, hence we set $H_\mathrm{ad}=H_\mathrm{ec}$. For the ionic conductivity of Nafion membranes, a power law from percolation theory with Arrhenius temperature correction was found to best fit various experimental data \cite{weber:04}:
\begin{multline}
\label{eq:sigma_p}
\sigma_\mathrm{p} = \epsilon_\mathrm{i}^{1.5}\,116\,\frac{\mathrm{S}}{\mathrm{m}}\max\{0,f-0.06\}^{1.5}\\
\times\exp\left[\frac{15\,\mathrm{kJ/mol}}{R}\left(\frac{1}{T_\mathrm{ref}}-\frac{1}{T}\right)\right]
\end{multline}
where $T_\mathrm{ref}=80\,^\circ\mathrm{C}$ and
\begin{equation}
f = \frac{\lambda V_\mathrm{w}}{\lambda V_\mathrm{w}+V_\mathrm{m}}
\end{equation}
denotes the volume fraction of water in the ionomer. $V_\mathrm{m}=1020/1.97\,\mathrm{cm}^3/\mathrm{mol}$ is the equivalent volume of the dry membrane (EW \cite{peron:10} divided by mass density \cite{nafion:16}) and $V_\mathrm{w}=18/0.978\,\mathrm{cm}^3/\mathrm{mol}$ the molar volume of liquid water at typical PEMFC operating conditions. In Eq.~\ref{eq:sigma_p} the Bruggeman correction $\epsilon_\mathrm{i}^{1.5}$ is used to account for the different ionomer contents $\epsilon_\mathrm{i}$ in the PEM and CLs \cite{dujc:18}.

Another crucial transport parameter is the water diffusivity in the ionomer $D_\lambda$. Experimental difficulties have led to large discrepancy in the literature data for Nafion \cite{kusoglu:17}. The measurements carried out by Mittelsteadt \& Staser \cite{mittelsteadt:11} appear to be among the most sophisticated. We refitted their data for Nafion membranes by a rational polynomial in $\lambda$ to obtain a smooth parameterization that captures all essential features of the data:
\begin{multline}
\label{eq:Dlambda}
D_\lambda = \epsilon_\mathrm{i}^{1.5} \frac{3.842\lambda^3-32.03\lambda^2+67.74\lambda}{\lambda^3-2.115\lambda^2-33.013\lambda+103.37}\,10^{-6}\frac{\mathrm{cm}^2}{\mathrm{s}}\\
\times\exp\left[\frac{20\,\mathrm{kJ/mol}}{R}\left(\frac{1}{T_\mathrm{ref}}-\frac{1}{T}\right)\right]
\end{multline}
This new fit for $D_\lambda$ is plotted in Fig.~\ref{fig:Dlambda}. Analogous to Eq.~\ref{eq:sigma_p}, the Bruggeman correction is used to model water diffusion through the partial ionomer content of the CLs.

\begin{Figure}
	\centering
	\captionsetup{type=figure}
	\includegraphics{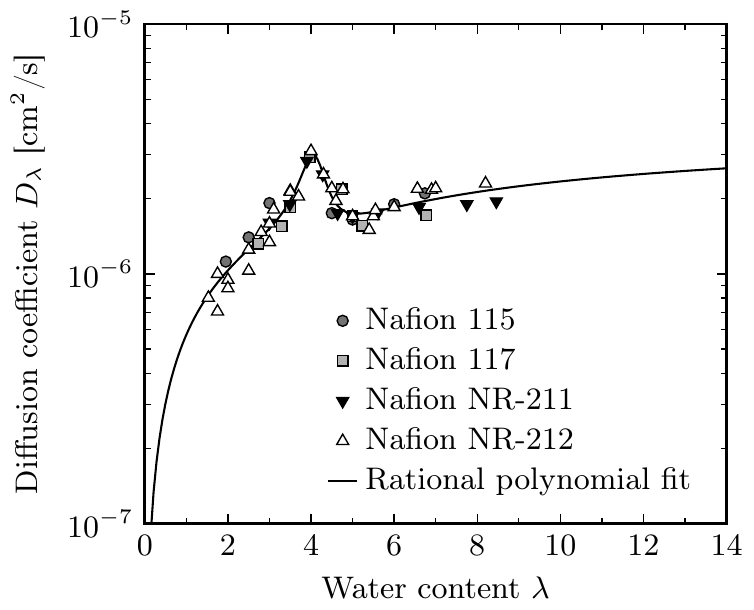}
	\captionof{figure}{Diffusion coefficient of water in Nafion membranes ($\epsilon_\mathrm{i}=1$) at $T=80\,^\circ\mathrm{C}$. Measurement data from ref.~\cite{mittelsteadt:11}.}
	\label{fig:Dlambda}
\end{Figure}

For the electro-osmotic drag coefficient, we adopt Springer's original linear law \cite{springer:91}
\begin{equation}
\label{eq:xi}
\xi = \frac{2.5\lambda}{22},
\end{equation}
and their sorption isotherm at $T=30\,^\circ\mathrm{C}$ is used to determine the equilibrium water content of the ionomer in Eq.~\ref{eq:Sad}:
\begin{equation}
\label{eq:sorption_isotherm}
\lambda_\mathrm{eq} = 0.043+17.81\mathrm{RH}-39.85\mathrm{RH}^2+36.0\mathrm{RH}^3
\end{equation}
where $\mathrm{RH}=x_{\mathrm{H}_2\mathrm{O}}/x_\mathrm{sat}$ is the relative gas humidity. A parameterization for the mass-transfer coefficients of water vapor in Nafion membranes has been determined by Ge et al.\ \cite{ge:05}:
\begin{equation}
\label{eq:kad}
k_\mathrm{a,d}=a_\mathrm{a,d}f\exp\left[\frac{20\,\mathrm{kJ/mol}}{R}\left(\frac{1}{T_\mathrm{ref}}-\frac{1}{T}\right)\right]
\end{equation}
where $a_\mathrm{a}=3.53\!\times\!10^{-3}\,\mathrm{cm/s}$ and $a_\mathrm{d}=1.42\!\times\!10^{-2}\,\mathrm{cm/s}$.

\subsection*{Transport in the porous media.}

To estimate the Fickean gas diffusivities within the pore space of the partially flooded CLs and GDLs, we amend the Chapman--Enskog formula \cite{bird:02} by the usual porous media correction factor for porosity and tortuosity, as well as by a saturation correction $(1-s)$ raised to the third power \cite{rosen:12}. For $X=\mathrm{H}_2,\mathrm{O}_2,\mathrm{H}_2\mathrm{O}$,
\begin{equation}
D_X = \frac{\epsilon_\mathrm{p}}{\tau^2}(1-s)^3D_{X,\mathrm{ref}}\left(\frac{T}{T_\mathrm{ref}}\right)^{1.5}\frac{P_\mathrm{ref}}{P}
\end{equation}
with the following prefactors:
\begin{itemize}
\item $D_{\mathrm{H}_2,\mathrm{ref}}=1.24\,\mathrm{cm}^2/\mathrm{s}$ (hydrogen in water vapor)
\item $D_{\mathrm{O}_2,\mathrm{ref}}=0.28\,\mathrm{cm}^2/\mathrm{s}$ (oxygen in air)
\item $D_{\mathrm{H}_2\mathrm{O},\mathrm{ref}}=1.24\,\mathrm{cm}^2/\mathrm{s}$ (water vapor in hydrogen, anode)
\item $D_{\mathrm{H}_2\mathrm{O},\mathrm{ref}}=0.36\,\mathrm{cm}^2/\mathrm{s}$ (water vapor in air, cathode)
\end{itemize}

As for the capillary pressure--saturation relationship, a large number of correlations have been suggested for various GDLs \cite{si:15}. We employ here the following expression that has been determined for Toray TGP-H-060 specifically \cite{nguyen:08}:
\begin{multline}
\label{eq:pc}
p_\mathrm{c}/\mathrm{Pa} = -0.00011\exp\left[-44.02(s-0.496)\right]\\
+278.3\exp\left[8.103(s-0.496)\right]-191.8
\end{multline}
The second parameter in the liquid water flux equation is the hydraulic permeability $\kappa$, which has a great impact on the liquid water distribution. It is modeled as \cite{nam:03}
\begin{equation}
\kappa = \left(10^{-6}+s_\mathrm{red}^3\right)\kappa_\mathrm{abs}
\end{equation}
where $\kappa_\mathrm{abs}$ denotes the absolute (intrinsic) permeability of the porous medium and the small numerical tolerance is added to avoid the singularity at $s_\mathrm{red}=0$.

A reasonable estimate for the liquid water saturation at the CGDL/GC interface $s_\mathrm{C}$ can be found with the notion of equivalent capillaries in the GDL through which the liquid water is transported. Using the Young-Laplace equation $p_\mathrm{c}=2\gamma\cos\theta/r$ with effective surface contact angle $\theta=130^\circ$ \cite{nguyen:08} and equivalent capillary radius $r=40\,$\textmu{}m \cite{holley:06} for Toray TGP-H-060, one finds an equivalent capillary surface pressure of $p_\mathrm{c}\approx2\,\mathrm{kPa}$. Using Eq.~\ref{eq:pc}, this translates to $s_\mathrm{C}\approx0.12$, which is the boundary value used here. Note that this does not pose any limitation on the boundary flux $j_s$ at the CGDL/GC interface, i.e., the interfacial liquid water flux will automatically be such that this pressure BC is met.

All remaining material properties are considered constant as listed in Tab.~\ref{tab:general_parameters}. The GDLs are assumed to be moderately compressed from 190 to 160\,\textmu{}m, corresponding to an applied clamping pressure of about 1.4\,MPa \cite{burheim:11}.

\section{Numerical implementation}

The model is implemented as a standalone MATLAB function !MMM1D! (short for one-dimensional Master MEA Model) which relies on standard built-in functionality only, for maximum compatibility. To solve the coupled equations, MATLAB's boundary value problem routine for ordinary differential equations !bvp4c! \cite{kierzenka:01} is used, a finite difference solver that implements the 3-stage Lobatto IIIa implicit Runge--Kutta method with automated mesh selection based on the residual, providing a 4\textsuperscript{th}-order accurate piecewise $C^1$-continuous solution. The return values of !MMM1D! are:
\begin{itemize}
\item !IU!, a two-column matrix containing the list of computed current densities and corresponding cell voltages (i.e., the polarization curve) of the fuel cell.
\item !SOL!, a cell array of solution structures returned by !bvp4c! (one for each row in !IU!) containing all potentials and fluxes as shown in Figs.~\ref{fig:potentials} and \ref{fig:fluxes}.
\end{itemize}
The complete source code as printed in the Appendix can be obtained from \url{https://www.isomorph.ch} for free. It is released under the 3-clause BSD license available from the same website, permitting unrestricted commercial and non-commercial use subject to the condition that the original source be referenced.

In the provided reference implementation, moderate !bvp4c! error tolerances are used (relative: $10^{-4}$, absolute: $10^{-6}$), resulting in an average of 54 mesh nodes used in total. The execution time for a full sweep over all cell voltages in steps of 50\,mV is a few seconds on a modern laptop computer, and the maximum absolute (relative) discretization error under these conditions is 4.4\,mA/cm$^2$ (0.23\%). Increased accuracy can be obtained by reducing these tolerance values if desired. All output plots shown in this paper have been obtained with very high accuracy using lower error tolerances (relative: $10^{-6}$, absolute: $10^{-10}$), which resulted in 158 mesh nodes on average for the base case.

\section{Simulation results}

\begin{table}
	\centering
	\caption{Operating conditions of the base case.\strut}
	\label{tab:operating_cond_base_case}
	\setlength{\tabcolsep}{0.3em}
	\begin{tabular}{llr}
	\toprule
	Symb. & Explanation & Value\\
	\midrule
	$P_\mathrm{A}$ & Gas pressure in anode gas channel & $1.5\,\mathrm{bar}$\\
	$P_\mathrm{C}$ & Gas pressure in cathode gas channel & $1.5\,\mathrm{bar}$\\
	$\mathrm{RH}_\mathrm{A}$ & Relative humidity in anode GC & $90\%$\\
	$\mathrm{RH}_\mathrm{C}$ & Relative humidity in cathode GC & $90\%$\\
	$s_\mathrm{C}$ & Liquid saturation at CGDL/GC interface & $0.12$\\
	$T_\mathrm{A}$ & Temperature of anode plate and GC & $70\,^\circ\mathrm{C}$\\
	$T_\mathrm{C}$ & Temperature of cathode plate and GC & $70\,^\circ\mathrm{C}$\\
	$\alpha_{\mathrm{H}_2}$ & Hydrogen mole fraction in dry fuel gas & $1.00$\\
	$\alpha_{\mathrm{O}_2}$ & Oxygen mole fraction in dry oxidant gas & $0.21$\\
	\bottomrule
	\end{tabular}
\end{table}

\subsection*{Base case.}

By default, our program code simulates a PEMFC at typical operating conditions referred to as the base case. These conditions are listed in Tab.~\ref{tab:operating_cond_base_case} and are used here to present the complete simulation output.

\begin{figure*}
	\centering
	\includegraphics{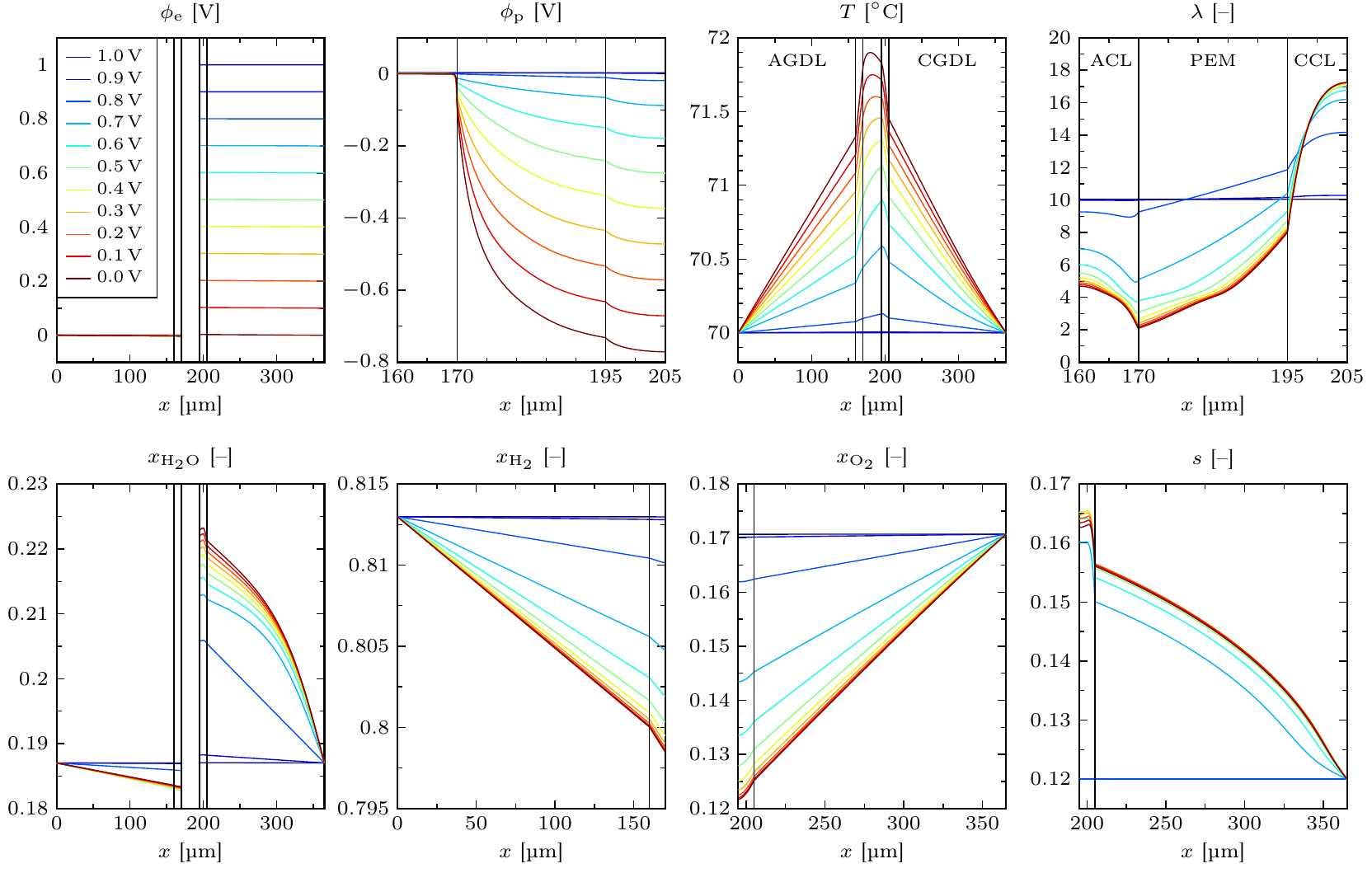}
	\caption{Through-plane potential profiles at different cell voltages for the base case. Subdomain boundaries are indicated by vertical lines. The anode is on the left, the cathode on the right hand side.}
	\label{fig:potentials}
\end{figure*}

\begin{figure*}
	\centering
	\includegraphics{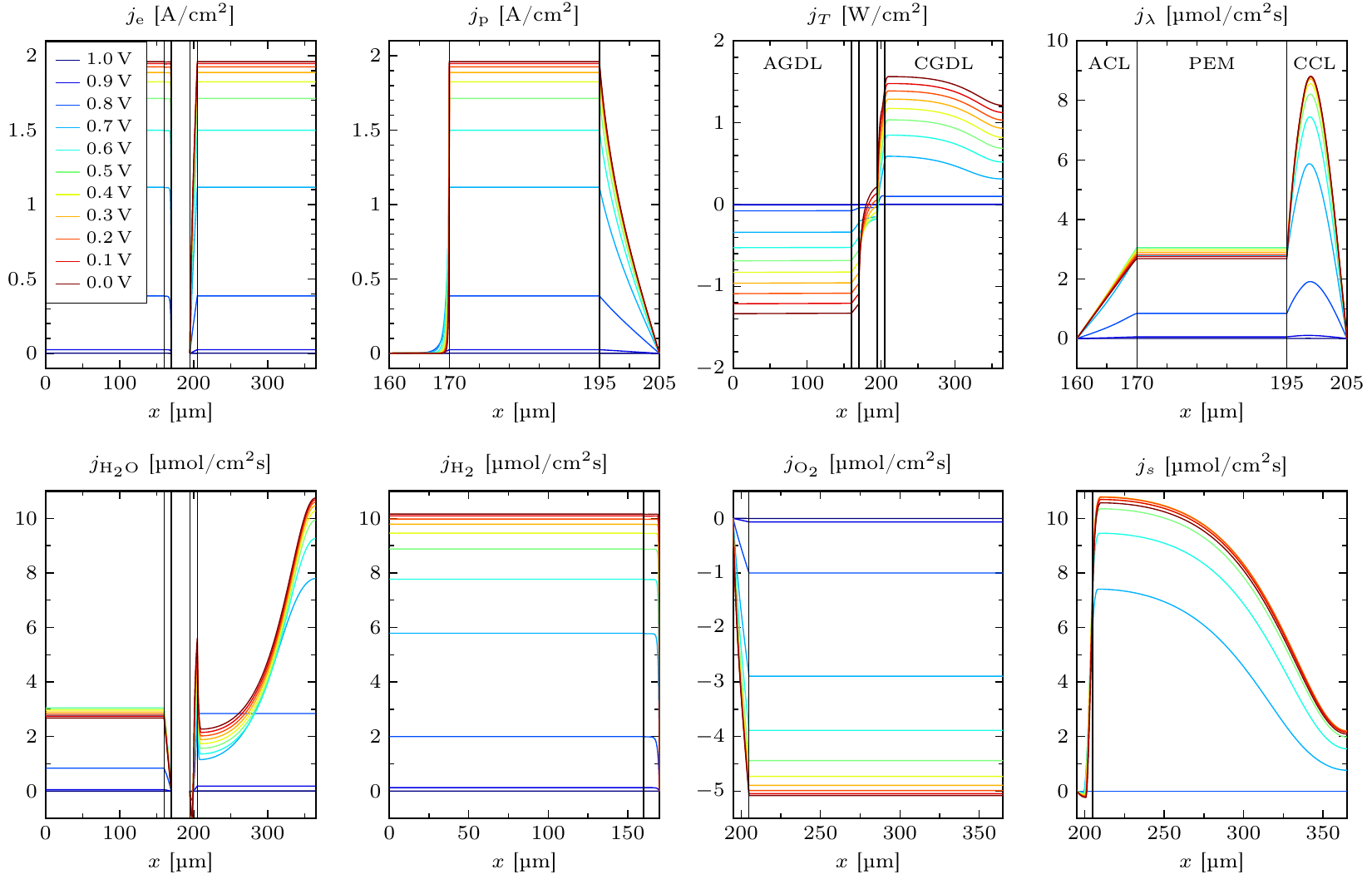}
	\caption{Through-plane flux profiles at different cell voltages for the base case.}
	\label{fig:fluxes}
\end{figure*}

The polarization curve is plotted in Fig.~\ref{fig:polarization_base_case} for the entire range of cell voltages in steps of 10\,mV.  Fig.~\ref{fig:potentials} shows the eight potentials and Fig.~\ref{fig:fluxes} the corresponding fluxes across the MEA layers as predicted by the model at different cell voltages in steps of 100\,mV. Each subplot is restricted to the support of the respective variable for a more detailed view where possible. All these plots are automatically generated by the MATLAB function.

The membrane phase potential and water content profiles deserve closer attention. Even though $S_\mathrm{p}\equiv0$ in the bulk membrane, $\phi_\mathrm{p}(x)$ exhibits significant curvature (Fig.~\ref{fig:potentials}). This is due to a strong spatial variation of the proton conductivity through the membrane, caused by a relatively steep decline of $\lambda$ toward the anode, which in turn is the result of strong electro-osmotic drag of dissolved water to the cathode. Parameterizations of the water diffusivity that predict larger values than Eq.~\ref{eq:Dlambda}, lower drag coefficients than Eq.~\ref{eq:xi}, and higher ionic conductivities than Eq.~\ref{eq:sigma_p} at low water content all yield higher $\lambda$ near the anode. This results in more flat potential profiles $\phi_\mathrm{p}(x)$ and consequently, higher current densities. These parameters are, in fact, among the material properties with the largest impact on the predicted fuel cell performance \cite{vetter:18}. An extensive study on this subject is currently underway at our institute.

\begin{table}
	\centering
	\caption{Key figures for the base case.\strut}
	\label{tab:key_figs_base_case}
	\begin{tabular}{lr}
	\toprule
	Quantity & Value\\
	\midrule
	Peak power density & 0.901\,W/cm$^2$\\
	Limit current density & 1.960\,A/cm$^2$\\
	Cell voltage $U$ at $I=1\,\mathrm{A/cm}^2$ & 0.720\,V\\
	Current density $I$ at $U=0.6\,\mathrm{V}$ & 1.499\,A/cm$^2$\\
	Peak temperature at $0.6\,\mathrm{V}$ & 70.90\,$^\circ$C\\
	Average temperature $\overline{T}$ at $0.6\,\mathrm{V}$ & 70.36\,$^\circ$C\\
	Minimum water content $\lambda$ at $0.6\,\mathrm{V}$ & 3.72\\
	Average water content $\overline{\lambda}$ at $0.6\,\mathrm{V}$ & 6.68\\
	Water flux through PEM at $0.6\,\mathrm{V}$ & 3.05\,\textmu{}mol/cm$^2$s\\
	Membrane resistance $R_\mathrm{PEM}$ at $0.6\,\mathrm{V}$ & 83.9\,m$\Omega$\,cm$^2$\\
	\bottomrule
	\end{tabular}
\end{table}

To further extend the data basis for future model comparison, we report on a few additional model characteristics for the base case, derived from the model output shown above. Tab.~\ref{tab:key_figs_base_case} lists some key figures, among which are the ohmic membrane resistance (excluding the contribution from the ionomer in the CLs)
\begin{equation}
R_\mathrm{PEM} = \int_\mathrm{PEM}\frac{1}{\sigma_\mathrm{p}}\diff x,
\end{equation}
the average MEA temperature
\begin{equation}
\overline{T} = \int_\mathrm{MEA}T\diff x \bigg/ \int_\mathrm{MEA}\diff x
\end{equation}
and the mean water content of the ionomer
\begin{equation}
\overline{\lambda} = \int_\mathrm{CCM}\epsilon_\mathrm{i}\lambda\diff x \bigg/ \int_\mathrm{CCM}\epsilon_\mathrm{i}\diff x.
\end{equation}
Here, the integration runs over the whole catalyst-coated membrane ($\text{CCM}=\text{ACL}\cup\text{PEM}\cup\text{CCL}$). These quantities are evaluated at $U=0.6\,\mathrm{V}$, but it is straightforward to use our program to calculate them at any other operating point.

\begin{Figure}
	\centering
	\captionsetup{type=figure}
	\includegraphics{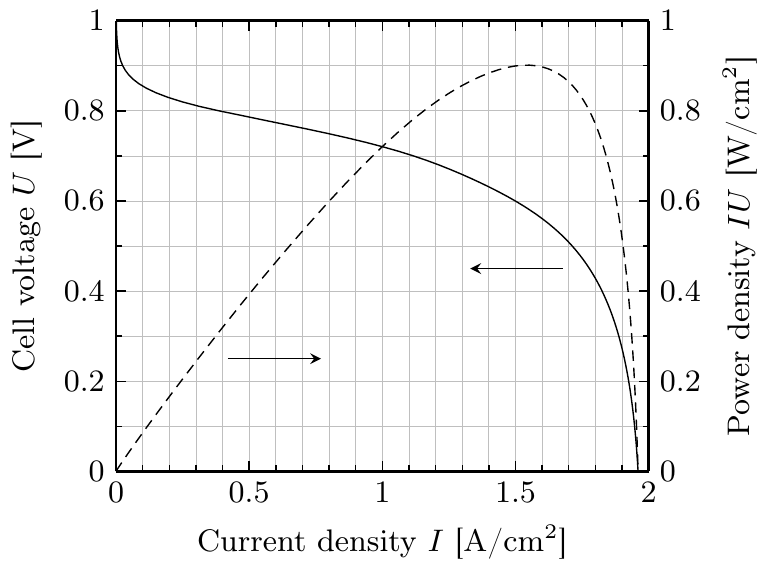}
	\captionof{figure}{Polarization curve of the base case.}
	\label{fig:polarization_base_case}
\end{Figure}

\begin{table*}
	\def\cc{\cellcolor{black!10}}
	\centering
	\caption{Stress tests recommended by the JRC. Shaded cells indicate deviations from the reference conditions.\strut}
	\label{tab:stress_tests}
	\begin{tabular}{lccccccccc}
	\toprule
	Input parameter & Unit & Reference & Test T1 & Test T2 & Test T3 & Test T4 & Test T5 & Test T6 & Test T7\\
	\midrule
	$P_\mathrm{A}$ & bar & 2.5 & 2.5 & 2.5 & 2.5 & 2.5 & 2.5 & \cc 1.6 & \cc 3.0\\
	$P_\mathrm{C}$ & bar & 2.3 & 2.3 & 2.3 & 2.3 & 2.3 & 2.3 & \cc 1.4 & \cc 2.8\\
	$\mathrm{RH}_\mathrm{A}$ & \% & 50 & \cc 85 & \cc 25 & 50 & \cc 25 & 50 & 50 & 50\\
	$\mathrm{RH}_\mathrm{C}$ & \% & 30 & \cc 85 & \cc 20 & \cc 20 & \cc 45 & \cc 45 & 30 & 30\\
	$T_\mathrm{A}=T_\mathrm{C}$ & $^\circ$C & 80 & \cc 45 & \cc 95 & \cc 95 & \cc 95 & \cc 95 & 80 & 80\\
	$s_\mathrm{C}^*$ & --- & 0 & \cc 0.12 & 0 & 0 & 0 & 0 & 0 & 0\\
	\midrule[\heavyrulewidth]
	Output parameter & Unit & Reference & Test T1 & Test T2 & Test T3 & Test T4 & Test T5 & Test T6 & Test T7\\
	\midrule
	$U$ at $0.1\,\mathrm{A/cm}^2$ & V & 0.829 & 0.863 & 0.789 & 0.822 & 0.837 & 0.848 & 0.816 & 0.834\\
	$U$ at $0.8\,\mathrm{A/cm}^2$ & V & 0.412 & 0.661 & --- & 0.435 & 0.531 & 0.605 & 0.359 & 0.435\\
	$I$ at $0.4\,\mathrm{V}$ & A/cm$^2$ & 0.809 & 0.991 & 0.556 & 0.842 & 0.960 & 1.137 & 0.770 & 0.826\\
	\midrule[\heavyrulewidth]
	\multicolumn{10}{l}{$^*$Parameter not part of JRC test specifications but set to match observed relative humidity ($s_\mathrm{C}=0$ if $\mathrm{RH}<1$).}\\
	\end{tabular}
\end{table*}

\begin{Figure}
	\centering
	\captionsetup{type=figure}
	\includegraphics{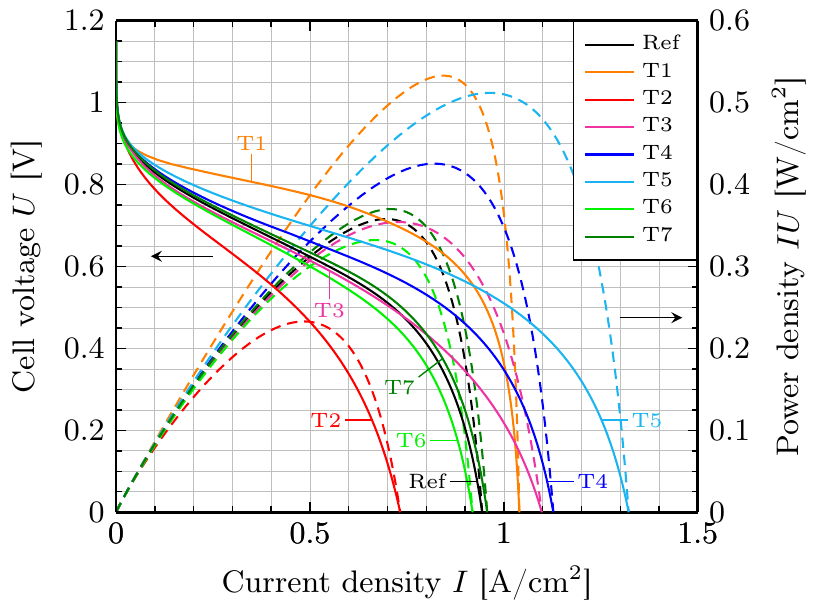}
	\captionof{figure}{Polarization curves of the stress tests.}
	\label{fig:polarization_stress_tests}
\end{Figure}

\begin{Figure}
	\centering
	\captionsetup{type=figure}
	\includegraphics{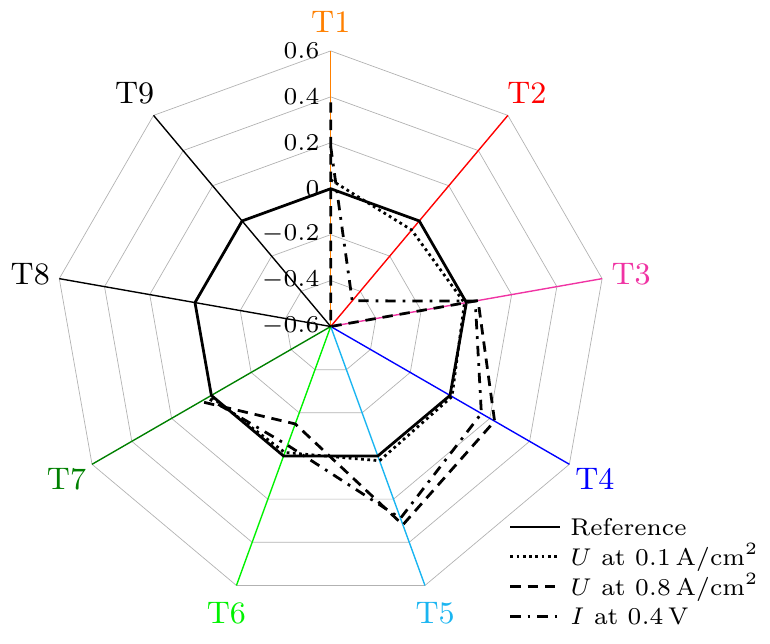}
	\captionof{figure}{Normalized cell voltages and current densities for the conducted stress tests recommended by the JRC.}
	\label{fig:spiderplots_stress_tests}
\end{Figure}

\subsection*{EU harmonized stress tests.}

A common limitation of fuel cell research in science and industrial applications to date is the lack of international standardization for experimental characterization and numerical modeling. To this end the Joint Research Centre (JRC) of the European Commission has recently issued a set of single-cell stress tests \cite{tsotridis:15}, termed T1 to T9 here, of which the first seven are directly applicable to the present 1D model. They define operating points aimed at characterizing the performance of a PEMFC under automotive conditions with variations in temperature (T1 and T2), gas humidification (T3 to T5) and gas pressure (T6 and T7), as detailed in Tab.~\ref{tab:stress_tests}. In order to establish a new baseline for harmonized fuel cell model comparison, validation and benchmarking, we subject our model to this series of stress tests and report the normalized output in the form recommended by the JRC in this section. These results may serve for comparison purposes in future PEMFC model development efforts.

The performance criteria agreed upon by the JRC specify three points on the polarization curve to be evaluated for each stress test. Tab.~\ref{tab:stress_tests} lists the model output data for the reference case and the applicable stress tests. The full resulting polarization curves are juxtaposed in Fig.~\ref{fig:polarization_stress_tests}. As proposed by the JRC, the three measured values are normalized by the corresponding values at reference conditions according to
\begin{equation}
\textrm{normalized result} = 1-\frac{\textrm{reference result}}{\textrm{stress test result}}
\end{equation}
and visualized as a spider plot in Fig.~\ref{fig:spiderplots_stress_tests}. Since the limit current density of T2 is only 0.733\,A/cm$^2$, the cell voltage of T2 at 0.8\,A/cm$^2$ is omitted from this compilation. Only Tests T1--T7 were conducted, because T8 and T9 (stoichiometry variation) are inapplicable to the present 1D model.

When interpreting the stress test results, it is important to note that a 1D through-plane MEA model simulates the characteristics of a differential fuel cell, i.e., one with a very small active cell area. The JRC specifies gas inlet conditions irrespective of gas channel length or cell area. In fuel cells with sufficiently long channels, the supplied gas gets humidified significantly as it flows downstream in the flow channels, making the area-averaged relative humidity much larger than at the inlet. With RH values between 20 and 85\%, the JRC test specifications therefore represent very dry conditions when applied to a differential cell. In order to mimic typical area-averaged conditions in the base case, the relative humidities were set to larger values for the default simulation setup (see Tab.~\ref{tab:operating_cond_base_case}), whereas we strictly abide by the dry JRC specifications here. Only T1 is humid enough to allow for the presence of liquid water in a differential cell as predicted by the model. All other stress tests are too dry for the RH to reach 100\% within the MEA of a differential cell. We therefore set $s_\mathrm{C}=0$ for all stress tests but T1.

The performance results shown in Fig.~\ref{fig:polarization_stress_tests} are thus strongly moisture-limited. T1, being the most humid scenario, allows for the best performance at moderate current densities, before electro-osmotic drag dries out the anode end of the membrane and the low temperature of only 45$^\circ$C becomes the limiting factor in the ionic conductivity (cf.\ Eq.~\ref{eq:sigma_p}), allowing the drier but hotter T3--T5 to perform better at low cell voltages. T2 (``dry/hot desert'') yields the worst performance prediction due to the extremely dry membrane ($\lambda\approx2.5$ on average). Owing to the stronger back diffusion of dissolved water in the ionomer, T4 (dry fuel) slightly outperforms T3 (dry air) in spite of the better ACL humidification under T3. Finally, we note that the dry test conditions of T6 and T7 do not allow the differential fuel cell to reach the large current density regime where a lot of fuel or oxidant is consumed and starvation becomes an issue. Under these test conditions, the model therefore exhibits very little sensitivity to pressure variations. This characteristic changes when less permeable (thicker) diffusion media and more humid gases are employed, which can easily be verified by using the provided MATLAB program.

\section{Conclusion}

With our free open MATLAB implementation of a macro-homogeneous two-phase PEMFC model, anyone with access to a MATLAB installation can readily run a state-of-the-art PEMFC simulation, substitute material parameterizations, add new model features or conduct parameter studies. It offers several major advantages over existing open-source codes. With less than 400 lines of compact, commented MATLAB code, it is exceptionally easy to read, maintain, extend and embed in other programs, with no third-party software, compilation of source code or knowledge of Linux and C++ or Python required. Both commercial and non-commercial use are permitted thanks to the BSD-like licence under which the program is published. We have provided the complete simulation output of the model at typical operating conditions, laying a new cornerstone in the ongoing effort of making numerical PEMFC modeling more transparent and accessible.

Despite these evident strengths, our model---like every model developed thus far---has limitations that need to be kept in mind. For the sake of simplicity, we intentionally neglect several chemical and physical processes occurring in real fuel cells that might become relevant under certain operating conditions. Among others certainly, the model does not account for the effects of heat convection, gas convection (due to a non-zero gas pressure gradient), gas species permeation and pressure-driven hydraulic permeation of water through the membrane, thermo-osmosis, Schroeder's paradox \cite{choi:03}, Knudsen diffusion, electrical and thermal contact resistance \cite{vetter:17}, mechanical deformation and other effects of clamping pressure, non-uniformity in material properties such as wettability and porosity, double layer effects, multi-step reaction kinetics, platinum oxide formation and the change of Tafel slope \cite{weber:14}, non-uniformity of ionic concentrations, ionic species migration, water droplet formation and detachment at the GDL/GC interface, the short-range effect observed in thin porous layers \cite{holzer:17a,holzer:17b}, degradation, unsteady phenomena, gravity, ice formation and melting, etc.

Many of the parameters and transport coefficients adopted for the present MEA model are subject to relatively wide variation in the literature and between different materials, while having a significant impact on the simulation results at the same time. Perhaps the largest source of uncertainty in the model lies in the liquid water flux through the porous media and across the interfaces. These are indeed modeling aspects for which no satisfactory universal solutions exist to date \cite{weber:14}. While PEMFC model development is still an ongoing process, an accessible numerical tool with which new parameterizations, interface conditions etc.\ can easily and quickly be tested, compared or validated against measurement data, can be key to further progress in this direction. Our open reference implementation of a 1D MEA model meets these requirements. A demonstration of how it can be utilized to quickly assess different material parameterizations has recently been presented \cite{vetter:18}. With a runtime of about a second on an ordinary laptop computer for a single simulation, it is suited even for time-critical applications. In cases where even less resources are available, it is possible to simplify the model in a number of ways, such as omitting the explicit account for liquid water through artificial extrapolation of Eq.~\ref{eq:sorption_isotherm} to the supersaturation regime as in ref.~\cite{springer:91}, merging the two diffusion equations for hydrogen and oxygen into one (because they are solved on disjunct subdomains), or removing the gas transport equations altogether. Moreover, as the first plot in Fig.~\ref{fig:potentials} shows, the electron phase potential $\phi_\mathrm{e}$ varies only little through the cell depth in the simulated base case. An order-of-magnitude analysis shows that for a GDL with thickness $L\sim\mathcal{O}(100\,$\textmu{}m$)$ and electric conductivity $\sigma_\mathrm{e}\sim\mathcal{O}(10^3\,\mathrm{S/m})$, the voltage loss associated with a current density of $I\sim\mathcal{O}(1\,\mathrm{A/m}^2)$ going through it is $IL/\sigma_\mathrm{e}\sim\mathcal{O}(1\,\mathrm{mV})$. In cases where voltage drops in this order of magnitude and the corresponding ohmic losses of $I^2L/\sigma_\mathrm{e}\sim\mathcal{O}(1\,\mathrm{mW/cm}^2)$ are deemed insignificant, the potential $\phi_\mathrm{e}$ may be replaced by constants in the GDLs. Doing so also in the CLs, on the other hand, would violate charge conservation ($S_\mathrm{e}=-S_\mathrm{p}$, cf.\ Tab.~\ref{tab:sources}) and lead to convergence difficulties.

Conversely, there is also much room for model extensions. Aside from the inclusion of the above-mentioned neglected effects, the model can be augmented by adding additional subdomains to represent MPLs, by using the Brinkman equation in place of Darcy's law, the Maxwell--Stefan equations in place of Fick's law for gas diffusion, the Nernst--Planck equation in place of Ohm's law, etc. Moreover, it is straightforward to add liquid water also on the anode side if desired. It is also possible to deeply refine the model in terms of its parameterization, for instance by including temperature dependence in the water sorption isotherm of the ionomer or in the electro-osmotic drag coefficient. Additional model complexity, detailed material property parameterizations and higher-dimensional models are being developed at our institute and are available upon request. More information can be found at \url{https://www.isomorph.ch}.

\section*{Acknowledgements}

Financial support from the Swiss National Science Foundation under the National Research Programme ``Energy Turnaround'' (NRP 70), project no.\ 153790, grant no.\ 407040\_153790, from the Swiss Commission for Technology and Innovation under contract no.\ KTI.2014.0115, through the Swiss Competence Center for Energy Research (SCCER Mobility), and from the Swiss Federal Office of Energy is gratefully acknowledged.

\newpage
\section*{Nomenclature}

\xentrystretch{-0.12} 

\begin{xtabular}{@{}ll@{}}
$a$ & Active surface area density [1/m]\\
$a_\mathrm{a}$ & Prefactor in $k_\mathrm{a}$ [m/s]\\
$a_\mathrm{d}$ & Prefactor in $k_\mathrm{d}$ [m/s]\\
$a_\mathrm{lg}$ & Liquid--gas interfacial area density prefactor [1/m]\\
$C$ & Total interstitial gas concentration [mol/m$^3$]\\
$D_X$ & Fickean diffusion coefficient of gas $X$ [m$^2$/s]\\
$D_{X,\mathrm{ref}}$ & Diffusivity of $X$ at reference conditions [m$^2$/s]\\
$D_\lambda$ & Diffusion coefficient of dissolved water [m$^2$/s]\\
$F$ & Faraday constant (96\,485.333\,C/mol)\\
$f$ & Water volume fraction in ionomer [--]\\
$\Delta G$ & Gibbs free energy difference [J/mol]\\
$\Delta H$ & Enthalpy of formation of liquid water [J/mol]\\
$H_\mathrm{ad}$ & Water ab-/desorption enthalpy [J/mol]\\
$H_\mathrm{ec}$ & Evaporation/condensation enthalpy [J/mol]\\
$I$ & Cell current density [A/m$^2$]\\
$i$ & Electrochemical reaction rate [A/m$^3$]\\
$i_0$ & Exchange current density [A/m$^2$]\\
$j_\mathrm{e}$ & Electronic flux [A/m$^2$]\\
$j_\mathrm{p}$ & Protonic flux [A/m$^2$]\\
$j_T$ & Heat flux [W/m$^2$]\\
$j_\lambda$ & Flux of dissolved water [mol/m$^2$s]\\
$j_X$ & Flux of gas $X$ [mol/m$^2$s]\\
$j_\mathrm{s}$ & Liquid water flux [mol/m$^2$s]\\
$k$ & Thermal conductivity [W/m\,K]\\
$k_\mathrm{a}$ & Water absorption transfer coefficient [m/s]\\
$k_\mathrm{c}$ & Water condensation transfer coefficient [m/s]\\
$k_\mathrm{d}$ & Water desorption transfer coefficient [m/s]\\
$k_\mathrm{e}$ & Water evaporation transfer coefficient [m/s]\\
$L$ & Layer thickness [m]\\
$M_\mathrm{w}$ & Molar mass of water [kg/mol]\\
$n$ & Interfacial unit normal vector [--]\\
$P$ & Absolute gas pressure [Pa]\\
$P_\mathrm{A}$ & Gas pressure in anode gas channel [Pa]\\
$P_\mathrm{C}$ & Gas pressure in cathode gas channel [Pa]\\
$P_\mathrm{ref}$ & Reference pressure (1\,atm, 101\,325\,Pa)\\
$P_\mathrm{sat}$ & Saturation water vapor pressure [Pa]\\
$p_\mathrm{c}$ & Capillary pressure [Pa]\\
$p_X$ & Partial pressure of gas $X$ [Pa]\\
$R$ & Gas constant (8.31446\,J/mol\,K)\\
$R_\mathrm{PEM}$ & Membrane resistance [$\Omega$\,m$^2$]\\
$r$ & Equivalent capillary radius [m]\\
$\mathrm{RH}$ & Relative gas humidity [--]\\
$\mathrm{RH}_\mathrm{A}$ & Relative humidity in anode gas channel [--]\\
$\mathrm{RH}_\mathrm{C}$ & Relative humidity in cathode gas channel [--]\\
$s$ & Liquid water saturation [--]\\
$s_\mathrm{C}$ & Saturation at cathode GDL/GC interface [--]\\
$s_\mathrm{im}$ & Immobile liquid water saturation [--]\\
$s_\mathrm{red}$ & Reduced liquid water saturation [--]\\
$S_\mathrm{F}$ & Substantial reaction rate [mol/m$^3$s]\\
$S_\mathrm{e}$ & Electron reaction rate [A/m$^3$]\\
$S_\mathrm{p}$ & Proton reaction rate [A/m$^3$]\\
$S_T$ & Heat source [W/m$^3$]\\
$S_{T,\mathrm{e}}$ & Joule heat source of electrons [W/m$^3$]\\
$S_{T,\mathrm{p}}$ & Joule heat source of protons [W/m$^3$]\\
$S_{T,\mathrm{r}}$ & Reaction heat source [W/m$^3$]\\
$S_{T,\mathrm{ad}}$ & Water ab-/desorption heat source [W/m$^3$]\\
$S_{T,\mathrm{ec}}$ & Evaporation/condensation heat source [W/m$^3$]\\
$S_\lambda$ & Dissolved water reaction rate [mol/m$^3$s]\\
$S_X$ & Reaction rate of gas $X$ [mol/m$^3$s]\\
$S_\mathrm{s}$ & Liquid water reaction rate [mol/m$^3$s]\\
$S_\mathrm{ad}$ & Water ab-/desorption source [mol/m$^3$s]\\
$S_\mathrm{ec}$ & Evaporation/condensation source [mol/m$^3$s]\\
$\Delta S$ & Reaction entropy [J/mol\,K]\\
$\Delta S_\mathrm{HOR}$ & Hydrogen oxidation reaction entropy [J/mol\,K]\\
$\Delta S_\mathrm{ORR}$ & Oxygen reduction reaction entropy [J/mol\,K]\\
$T$ & Absolute temperature [K]\\
$T_\mathrm{A}$ & Temperature of anode plate and GC [K]\\
$T_\mathrm{C}$ & Temperature of cathode plate and GC [K]\\
$T_\mathrm{ref}$ & Reference temperature (80\,$^\circ$C, 353.15\,K)\\
$\overline{T}$ & Mean MEA temperature [K]\\
$U$ & Cell voltage [V]\\
$V_\mathrm{m}$ & Acid equivalent volume of membrane [m$^3$/mol]\\
$V_\mathrm{w}$ & Molar volume of liquid water [m$^3$/mol]\\
$x$ & Through-plane coordinate [m]\\
$x_X$ & Mole fraction of gas $X$ [--]\\
$x_{\mathrm{H}_2\mathrm{O}}^\mathrm{A}$ & Water vapor mole fraction in anode GC [--]\\
$x_{\mathrm{H}_2\mathrm{O}}^\mathrm{C}$ & Water vapor mole fraction in cathode GC [--]\\
$x_{\mathrm{H}_2}^\mathrm{A}$ & Hydrogen mole fraction in anode GC [--]\\
$x_{\mathrm{O}_2}^\mathrm{C}$ & Oxygen mole fraction in cathode GC [--]\\
$x_\mathrm{sat}$ & Saturation water vapor mole fraction [--]\\
$\alpha_{\mathrm{H}_2}$ & Mole fraction of hydrogen in dry fuel gas [--]\\
$\alpha_{\mathrm{O}_2}$ & Mole fraction of oxygen in dry oxidant gas [--]\\
$\beta$ & Half-reaction symmetry factor [--]\\
$\gamma$ & Surface tension of water [N/m]\\
$\gamma_\mathrm{c}$ & Water condensation rate [1/s]\\
$\gamma_\mathrm{e}$ & Water evaporation rate [1/s]\\
$\epsilon_\mathrm{i}$ & Ionomer volume fraction [--]\\
$\epsilon_\mathrm{p}$ & Pore space volume fraction (porosity) [--]\\
$\eta$ & Activation overpotential [V]\\
$\theta$ & Effective contact angle [deg]\\
$\kappa$ & Hydraulic permeability [m$^2$]\\
$\kappa_\mathrm{abs}$ & Absolute (intrinsic) permeability [m$^2$]\\
$\lambda$ & Ionomer water content [--]\\
$\lambda_\mathrm{eq}$ & Equilibrium ionomer water content [--]\\
$\overline{\lambda}$ & Mean ionomer water content [--]\\
$\mu$ & Dynamic viscosity of liquid water [Pa\,s]\\
$\xi$ & Electro-osmotic drag coefficient [--]\\
$\sigma_\mathrm{e}$ & Electric conductivity [S/m]\\
$\sigma_\mathrm{p}$ & Protonic conductivity [S/m]\\
$\tau$ & Pore tortuosity [--]\\
$\phi_\mathrm{e}$ & Electrode phase potential [V]\\
$\phi_\mathrm{p}$ & Electrolyte phase potential [V]\\
$\Delta\phi$ & Galvani potential difference [V]\\
$\Delta\phi_0$ & Reversible potential difference [V]\\
\end{xtabular}

\end{document}